\documentclass[prb,twocolumn,superscriptaddress]{revtex4}
\usepackage{graphicx}
\begin{document}

\title[Phonons in MgB$_2$]{ Phonons in MgB$_2$  by
 Polarized Raman Scattering on Single Crystals}

\date{\today}

\author{J. Hlinka}
\author{ I. Gregora}
\author{ J. Pokorn\'{y}}
\affiliation{Institute of Physics ASCR, Praha, Czech Republic}
\author{Plecenik}
\affiliation{Institute of Electrical Engineering SAS, Bratislava,
Slovak Republic}
\author{P. K\'{u}\v{s}}
\affiliation{Department of Solid State Physics, FMFI, Comenius
University, Bratislava, Slovak Republic}
\author{L. Satrapinsky}
\author{\v{S}. Be\v{n}a\v{c}ka}
\affiliation{Institute of Electrical Engineering SAS, Bratislava,
Slovak Republic}

\begin{abstract}
The paper presents detailed Raman scattering study of the
unusually broad E$_{2g}$ phonon mode in MgB$_2$ crystal. For the
first time, it is shown by the polarized Raman scattering on
few-micron-size crystallites with natural faces that the observed
broad Raman feature really does obey the selection rules of an
E$_{2g}$ mode. Raman spectra on high quality polycrystalline
superconducting MgB$_2$ wires reveal a very symmetric E$_{2g}$
phonon line near 615 cm$^{-1}$ with the room temperature linewidth
of 260 cm$^{-1}$ only. Additional scattering of different
polarization dependence, observed in certain crystallites is
interpreted as weighted phonon density of states induced by
lattice imperfections.
\end{abstract}

\pacs{74.25.Kc, 78.30.-j, 63.20.-e}

\maketitle

Recent discovery of superconductivity\cite{nature1} in the
hexagonal magnesium diboride (P6/mmm, Z=1) has motivated numerous
experimental and theoretical investigations. Among others, it was
claimed that the superconductivity in this compound arises due to
the strong electron-phonon
coupling.\cite{BCS,Kortus,Bela,An,BOH,Yil} This electron-phonon
coupling was found to be particularly strong for the E$_{2g}$
optic phonon mode.\cite{An,BOH,Yil} In fact, this E$_{2g}$ phonon
mode is the only first order Raman active mode in MgB$_2$, so that
Raman scattering is a very suitable tool for investigation of this
particular phonon mode.

Previous to our investigations, three Raman studies of MgB$_2$
were reported.\cite{BOH,GON,CHEN} The paper by Bohnen, Heid and
Henker\cite{BOH} reported Raman spectrum on a 10 micron size
crystalline grain from a commercially available powder. The
maximum of the observed broad asymmetric peak near 72~meV (580
cm$^{-1}$) corresponds rather well to their harmonic frequencies
calculated {\em ab-initio}. The unusually large width is ascribed
to the predicted strong electron phonon coupling. The second paper
by Goncharov et al.\cite{GON} is devoted to the evolution of the
Raman spectra with applied pressure. Apart from the unusually
large positive pressure shift of the frequency, it was found that
the Raman spectrum of individual micron size grains is essentially
identical with that of the powder. Ambient pressure spectra were
fitted by a Gaussian profile with maximum near 620 cm$^{-1}$ and a
width (FWHM) of 300 cm$^{-1}$.
 The origin of the  observed signal was questioned but it was
concluded that the most probable interpretation is that of first
order Raman response from the E$_{2g}$ phonon mode. A subsequent
paper by Chen at al.\cite{CHEN} shows the same broad peak (620
cm$^{-1}$) at temperatures of 15 and 45K in Raman spectra taken
from polycrystalline MgB$_2$ samples with 0.15-0.3 micron size
grains. They performed polarization analysis but no significant
difference between the shape and intensity of spectra measured
with parallel and antiparallel polarizers was observed, even when
using the Raman microprobe. Based on these findings and in
contrast to the two previous studies\cite{BOH,GON}, Chen at
al.\cite{CHEN} proposed that the observed broad feature can be
interpreted as a disorder induced phonon contribution throughout
the Brillouin zone, {\em i.e.} a kind of weighted phonon density
of states.

The present experiments were carried out using a Renishaw Raman
microscope with 514.5~nm (2.41~eV) argon laser excitation. The
instrument allows measurements of polarized Raman spectra in back
scattering from a spot size down to 1-2 microns in diameter. As a
polycrystalline material we have used MgB$_2$ wires with T$_{\rm
c}$=40~K, produced by sealing a boron fiber with a tungsten core
and magnesium chips (see Ref.~\onlinecite{PLE} for preparation
details and characterization).
\begin{figure}[!]
\hspace{0cm} \centerline{\includegraphics[width= 7cm, clip=true]
    {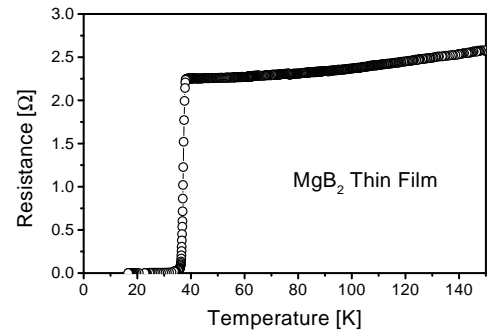}}
 \caption{Resistance versus temperature dependence of the
  MgB$_2$ thin film\label{Fig1}}
\end{figure}
Microscopic MgB$_2$ single crystals were located on the MgB$_2$
thin films synthetized by ex-situ annealing of boron thin films.
Thin boron films with nominal thickness from 100~nm to 200 ~nm
were first deposited on a non-heated, randomly oriented,
Al$_2$O$_3$ substrate from a Ta resistive heater in vacuum of
8$\times$10$^{-4}$ Pa. The thin boron film was packed together
with magnesium chips into a Nb tube. The Nb tube was inserted into
an annular furnace where Ar atmosphere of 3 kPa was maintained
during the whole annealing cycle. The furnace temperature was
increased from room temperature to 800 $^o$C in one hour, kept for
half an hour at 800 $^o$C and subsequently decreased to the room
temperature in 5 minutes. The resistance vs. temperature curves
(R-T, Fig.~\ref{Fig1}) show the onset of superconducting state
T$_{\rm con}$ near 38~K and zero resistance temperature of about
37~K (from 90~\% to 10~\%).

\begin{figure}
\hspace{0cm} \centerline{\includegraphics[width= 6cm, clip=true]
    {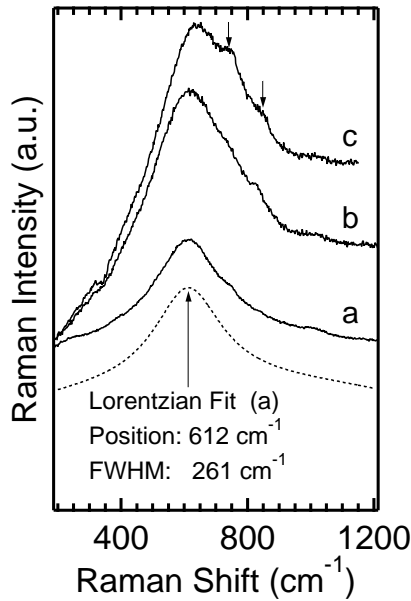}}
 \caption{Unpolarized back scattering Raman spectra. Solid curve
(a) for the polycrystalline MgB$_2$ wire; solid curves (b) and (c)
for hexagonal faces of two distinct crystallites; broken curve is
the Lorentzian fit of the curve (a). See text.\label{Fig2}}
\end{figure}
\begin{figure}
\hspace{0cm} \centerline{\includegraphics[width= 6cm, clip=true]
    {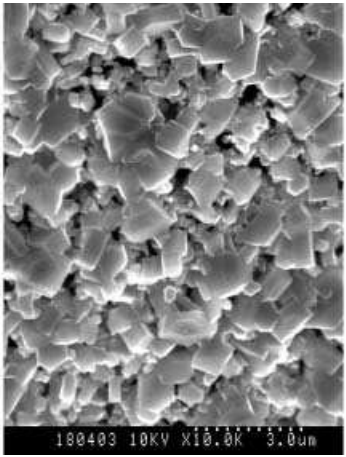}}
 \caption{The SEM micrograph of the MgB$_2$ thin film.
\label{Fig3}}
\end{figure}

\begin{figure}
\hspace{0cm} \centerline{\includegraphics[width= 6cm, clip=true]
   {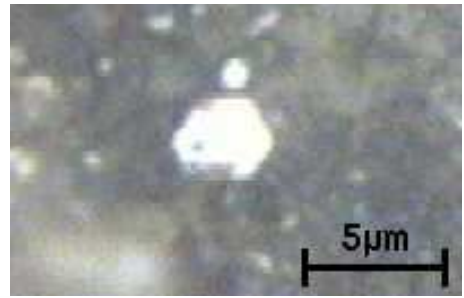}}
 \caption{Micrograph of a hexagonal MgB$_2$ crystallite on surface
 of a rough thin layer as seen in the optical microscope. Bright
 areas indicate
natural crystal faces parallel to the substrate.\label{Fig4}}
\end{figure}

\begin{figure}
\hspace{0cm} \centerline{\includegraphics[width= 6cm, clip=true]
   {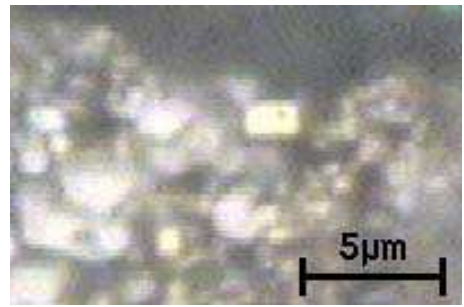}}
 \caption{Micrograph showing the rectangular MgB$_2$ crystallite
  in the optical microscope. See the preceding figure.\label{Fig5}}
\end{figure}

\begin{figure}
\hspace{0cm} \centerline{\includegraphics[width= 6cm, clip=true]
    {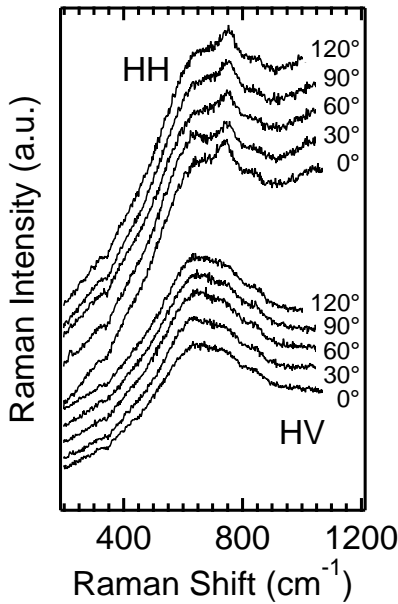}}
 \caption{Polarized Raman spectra taken
from the hexagonal face of the MgB$_2$ crystallite in
Fig.~\protect\ref{Fig4} as a function of the angle between its
horizontal edge and the incident polarization direction. The
spectra at zero angle have no vertical offset.\label{Fig6}}
\end{figure}

Typical unpolarized Raman spectra taken from polycrystalline wires
 (Fig.~\ref{Fig2}, spectrum a) show essentially the same broad
 feature as those reported
earlier\cite{BOH,GON,CHEN}. It can be fitted perfectly with a
Lorentzian profile at 612 cm$^{-1}$ with FWHM of 261 cm$^{-1}$. We
have noticed that our peak is somewhat sharper and more
symmetrical than that of the previously published spectra. The SEM
picture of thin films (Fig.~\ref{Fig3})
 exhibits polycrystalline morphology  with
single crystal diameter of about 1 $\mu$m. Distinct
 crystallites are oriented more or less randomly.
 Observation in an optical
microscope (reflected light) emphasizes the crystal faces which
are parallel to the substrate. Among them, several larger
hexagonal or rectangular crystal faces could be identified: two
such faces are shown in Figs.~\ref{Fig4} and \ref{Fig5}. We have
assumed that these faces correspond to the high symmetry planes of
MgB$_2$ crystallites, {\em i.e. } perpendicular (Fig.~\ref{Fig4})
and parallel (Fig.~\ref{Fig5}) to the hexagonal axis. The
unpolarized Raman spectra taken from hexagonal faces of two
different crystallites are shown in Fig.~\ref{Fig2} (spectra b and
c). The investigated broad peak is superimposed on a considerable
luminescence background of unknown origin, whose intensity differs
from crystallite to crystallite. Simultaneously, some finer
features are observed mainly on the high frequency side of the
investigated peak (see arrows on Fig.~\ref{Fig2}, spectrum c).
Figure~\ref{Fig6} shows that the "fine structure" is more
pronounced in the polarized Raman spectra with parallel polarizers
than in the cross-polarized spectra. The same holds for the
luminescence background. This effect, most probably caused by some
kind of defects, is discussed in more detail below.


\begin{figure}
\vspace{0cm} \centerline{\includegraphics[width= 6cm, clip=true]
    {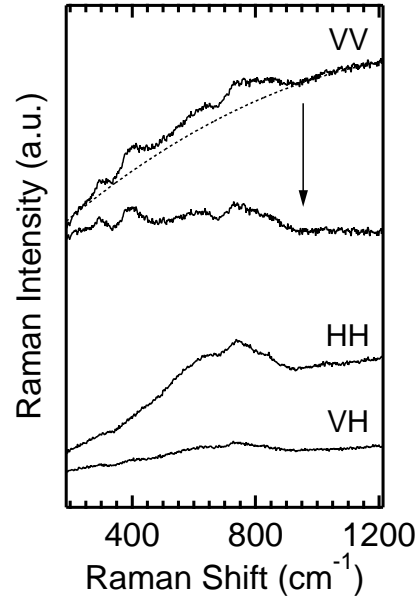}}
 \caption{Polarized Raman spectra taken
from the rectangular face of the MgB$_2$ crystallite in
Fig.~\protect\ref{Fig5}. The arrow shows difference between VV
spectrum and interpolated parabolic background due to the
luminescence (dashed curve).\label{Fig7}}
\end{figure}

Assuming that the broad peak corresponds to the doubly degenerate
E$_{2g}$ phonon of P6/mmm crystal, the standard Raman selection
rules require the allowed scattering to be independent of
polarization if both polarization directions are perpendicular to
the axis (back scattering from the hexagonal face). If at least
one of the polarization directions is parallel to the hexagonal
axis, the scattering is forbidden. We have found that the broad
peak response actually obeys these rules. Independence of the
rotation around the hexagonal axis for both parallel (HH) and
crossed (HV) polarization geometry is obvious from
Fig.~\ref{Fig6}. Moreover, the expected extinction of the broad
peak in measurements with one or both polarization directions
parallel to the hexagonal axis is evident in VV and VH spectra
shown in Fig.~\ref{Fig7}. Here the H and V refer to the horizontal
and vertical direction of the polarization of incident or
scattered photons with respect to the edges of the rectangular
crystalline face shown in Fig~\ref{Fig5}. We note that the tacit
assumption that the shorter edge is parallel to the hexagonal axis
was fulfilled for all of the investigated rectangular crystallites
and is further corroborated by our frequent observation of
generally oriented MgB$_{2}$ crystallites with hexagonal
plate-like habitus on the same substrate. In addition, the HH
spectra of Fig.~\ref{Fig6} and Fig.~\ref{Fig7} are very similar,
despite the fact that they are taken from different crystallites.
Our measurements thus confirm the standard selection rules of
E$_{2g}$ modes in 6/mmm point group for the main peak at 620
cm$^{-1}$.


 On the other hand, the additional "fine structure" does not follow
the same selection rules as the main peak, namely it does not
disappear in VV spectra shown in the Fig.~\ref{Fig7}. After
subtraction of an interpolated smooth parabolic luminescence
background, one can obtain a structure indicated by the arrow in
the Fig.~\ref{Fig7}, which somehow reminds of the phonon density
of states. It shows five humps at about 300, 400, 600, 750 and
830~cm$^{-1}$, which roughly correspond to the peaks on the
reported phonon density of states\cite{BOH,Yil,Osborn} or
Eliashberg function\cite{BOH}, or at least correspond to the
frequency range of the MgB$_{2}$ phonon spectrum. In fact, a
signature of these humps can be identified in most of the
displayed spectra (for example, the dip in between of the 300 and
400 cm$^{-1}$ humps is clearly marked in all the spectra shown in
Fig.~\ref{Fig6}). Thus, it is highly probable that this extra
Raman signal is related to disorder-induced contribution due to
relaxation of the wave vector selection rules. Such or similar
mechanism could perhaps also explain why the previous measurements
\cite{BOH,GON,CHEN} systematically provided asymmetric profiles,
steeper on the low-frequency side.

Just after completing our investigation, a new detailed paper on
polarized Raman study of 40x30x5 micron size MgB$_{2}$ crystallite
by Martinho et al.\cite{MAR} appeared in the cond-mat E-print
archive. Their surprising observation of complete breakdown of the
polarization selection rules of the investigated E$_{2g}$ mode (
at 630cm$^{-1}$, FWHM 275~cm$^{-1}$) is in flagrant disagreement
with our measurements, done with the same excitation wavelength.
On the other hand, Martinho et al.\cite{MAR} made important
observation that the phonon linewidth decreases with temperature
down to about 180~cm$^{-1}$ at T$=$15~K, which corroborates the
prediction\cite{Yil} of the giant anharmonicity of the E$_{2g}$
mode. Let us assume that the maximum of the observed E$_{2g}$
Raman line corresponds to the energy difference between the two
lowest energy levels in an anharmonic potential for the E$_{2g}$
mode coordinate of a same shape as calculated {\em ab initio} by
Yildirim et al\cite{Yil}. In this case the real harmonic frequency
corresponding to our best peak frequency of 612~cm$^{-1}$ (76~meV)
is merely about 443~cm$^{-1}$ (55~meV), which is quite close to
the {\em ab initio} values given by Kortus et al\cite{Kortus}
(470~cm$^{-1}$), Yildirim et al\cite{Yil} (60.3~meV) and not too
far from the values of Bohnen, Heid and Henker\cite{BOH} (66.5 and
70.8~meV for experimental and calculated geometries,
respectively).

In conclusion, we have presented a critical review of available
Raman studies on phonons in MgB$_{2}$ and completed it by more
detailed characterization. Our experiments suggest that the
observed Raman response consists of a broad but symmetrical phonon
E$_{2g}$ line near 615~cm$^{-1}$, complemented by a more
complicated weaker component reflecting phonon density of states
in crystals with defects. An extra component of this type might
also explain the asymmetry of the broad E$_{2g}$ line in the
previously published observations. Finally, the E$_{2g}$
assignment of the principal peak is directly proven by
polarization analysis of Raman scattering from oriented MgB$_{2}$
microcrystallites.

\begin{acknowledgements}
This work has been supported by the Slovak Grant Agency for
Science (Grants No. 2/7199/20 and 1/7072/20)
 and by the Grant Agency of the Czech
Republic (Postdoc project 202/99/D066).
\end{acknowledgements}

\end{document}